\documentclass[conference]{IEEEtran} 
\IEEEoverridecommandlockouts
\usepackage{cite}
\usepackage{amsmath,amssymb,amsfonts}
\usepackage{algorithmic}
\usepackage{graphicx}
\usepackage{textcomp}
\usepackage{xcolor}
\usepackage{svg}
\usepackage{tabularx}
\usepackage{booktabs}

\usepackage{siunitx}
\DeclareSIUnit{\GBs}{\frac{GB}s}

\def\BibTeX{{\rm B\kern-.05em{\sc i\kern-.025em b}\kern-.08em
    T\kern-.1667em\lower.7ex\hbox{E}\kern-.125emX}}
\begin{document}

\title{Evaluation Pipeline for systematically searching for Anomaly Detection Systems
}

\author{\IEEEauthorblockN{Florian Rokohl}
\IEEEauthorblockA{\textit{Institute of Applied Microelectronics}\\ \textit{and Computer Engineering} \\
\textit{University Rostock}\\
Rostock, Germany \\
0009-0001-6924-0585}
\and
\IEEEauthorblockN{Alexander Lehnert}        
\IEEEauthorblockA{\textit{Institute of Applied Microelectronics}\\ \textit{and Computer Engineering} \\
\textit{University Rostock}\\
Rostock, Germany \\
0000-0001-5111-4620}
\and
\IEEEauthorblockN{Marc Reichenbach}
\IEEEauthorblockA{\textit{Institute of Applied Microelectronics}\\ \textit{and Computer Engineering} \\
\textit{University Rostock}\\
Rostock, Germany \\
0000-0002-9687-6247}
\thanks{This work was funded by the German Federal Office for Information Security (BSI) under grant number 01MO23015D.}
}

\maketitle

\begin{abstract}
  Digitalization in the medical world provides major benefits while making it a target for attackers and thus hard to secure.
  To deal with network intruders we propose an anomaly detection system on hardware to detect malicious clients in real-time.
  We meet real-time and power restrictions using FPGAs.
  Overall system performance is achieved via the presented holistic system evaluation.
\end{abstract}

\begin{IEEEkeywords}
  Anomaly Detection, Network Sketches 
\end{IEEEkeywords}

\section{Introduction}

Digitalization in the medical world offers major benefits over traditional paper-based systems found in many current medical institutions.
Devices and professionals are interconnected, data is aggregated, which in turn enables fast evaluation and accurate diagnoses.
As such, there are extensive efforts to digitalize processes in medical environments.
On the flip side, this development opens new doors for intruders.
Big medical institutions handle large amounts of confidential data, making them a favorable target of digital attacks.

The introduction of wireless communication does not solve the security risks.
Rather it makes the network itself more vulnerable.
A plethora of devices also opens up the same number of possible targets.
Security for such networks can not be granted.
Therefore, we investigate the detection of intruders into these networks.

On average, there is a time window of 14 days between an attacker gaining access to the network and actual data leaks~\cite{shulman_forschung_2023}.
Meanwhile, an attacker emits different communication behavior to a conventional network client, especially in a medical environment.
Within this time window the \textbf{anomalous behavior of malicious clients} can be detected when investigating the network traffic.
Such an approach exceeds detecting single types of attacks, such as e.g.\ denial of service (DOS) attacks, but is capable of detecting anomalies in general.
In this ever changing landscape of attacks the optimal soution must be a reconfigurable while real-time capable one.
Therefore, we present a modular system architecture leveraging FPGAs.

The implementation of a generalized anomaly detection system is non-trivial.
Firstly, there exists a multitude of anomaly detection methods, with no best solution for any anomaly and each building upon different data aggregation.
Secondly, the performance and efficiency of these methods depends vastly on the hardware implementation.
Further, in large networks, as is the case in medical environments, anomaly detection on edge can be considered.
In this work, we propose a holistic approach to evaluating various combinations of i) anomaly detection methods on ii) hardware architectures iii) targeted at different requirements.
This evaluation not only considers the detection itself, but it also considers any type of preprocessing required.
The implementation using FPGAs enables our modular framework for data preprocessing and anomaly detection.
As a first step, we present the implementation of a modular feature extractor for real time analysis of network traffic.

\section{Current Research}

In this section, we present anomaly detection methods on recent network traffic using network sketches.
The current normal and global state determines whether a network package should be classified as an anomaly.
Network sketches hold an approximation of predefined relevant information of the current network state and thus enable this classification.
Network flow features are recoreded live using hash functions and buffer structures to aggregate the observed network traffic into a fixed-size memory.
Furthermore, network sketches enable reverting of IP addresses from the buffered hashes.
In addition, these sketches are capable of supporting high throughputs~\cite{tong_sketch_2018}, which in turn enables live detection of anomalies.
A high number of different implementations do exist which have there own use case~\cite{zhou_generalized_2019}.

Nassif et al.\ show present a review of various machine learning algorithms targeted at anomaly detection~\cite{nassif_machine_2021}.
These algorithms are categorized into flow-based and package-based based on the type of input data processed.
The flow-based algorithms is dealing with extracted network flow features, such as number of sent packages in a network flow.
This information is either given by a dataset or in collected at multiple end points inside the network and aggregated in time windows.
Yi et al.\ proposed a system consisting of LSTM and CNN with a network sketches as preprocessor~\cite{yi_network_2023}.
While such an approach is incapable of identifying singular anomalous network packages, they can classify if a whole network flow, described by the featues, is anomalous.
Such features can be for instance number of bits from source to destination, TCP base sequence number or the source inter-packet arrival time.
The other group of machine learning algorithms is dealing with raw package data~\cite{nassif_machine_2021}.
Sun et al.\ present a CNN processing network traffic as time-series to detect different abnormal behavior with high precision.~\cite{sun_deep_2021}
These models can be easily deployed but suffer from the high throughput of network packages in the real world.
While performance of such approaches is similar, flow-based methods are often preferred as they allow for high throughput implementations.

\section{Methods}

\begin{figure}
  \label{system_overview}
  \centering
  \includegraphics[height=0.26\textheight]{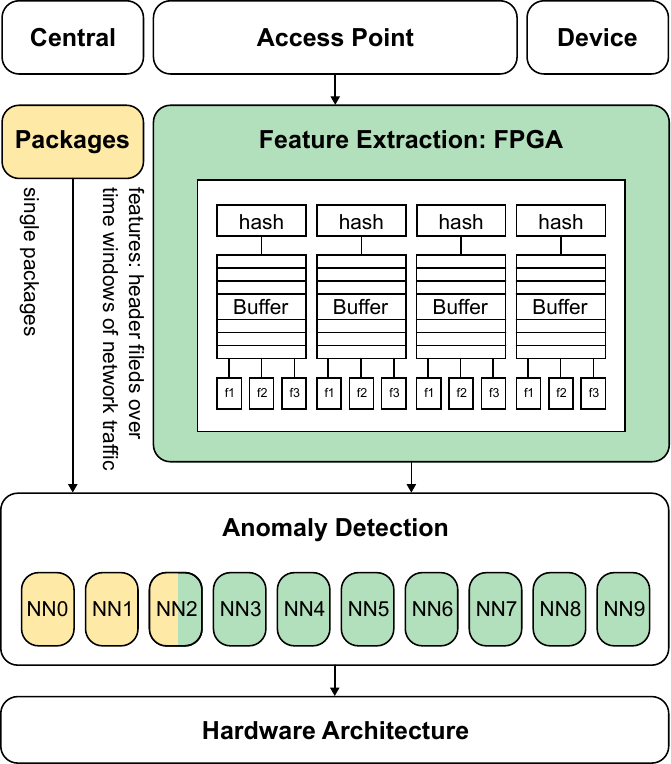}
  \caption{Evaluation Pipeline Overview}
\end{figure}

Our holistic evaluation appraoch is shown in Figure~\ref{system_overview}.
There are four dimensions to the evaluation: (i) the position within the network (ii) any required preprocessing (iii) anomaly detection methods and (iv) hardware architecture used for implementation.
For a complete system evaluation, all four levels have to be considered.

Depending on the neural network, various preprocessing steps, such as feature extraction, might be required.
The combination of feature extraction and the anomaly detection method is mapped to a hardware architecture in a second step.
In turn, each of these implemented methods is then evaluated for resource and power efficiency.
Further it is required, that any valid implementation is capable of the throughput induced by the 5G network, which poses as the data source.
The evaluation itself will be conducted on the basis of state-of-the-art network communication datasets.
These datasets must be representative for the application at hand, i.e. a medical network, as well as the place, where the system is enrolled.

Several methods of anomaly detection will be explored.
Such methods include the ones presented earlier~\cite{nassif_machine_2021, sun_deep_2021, yi_network_2023}, as well as novel approaches.
This way, a representative evaluation of methods of anomaly detection is done.
All solutions will not only be investigated in terms of resource cost, but also performance in terms of accuracy of detecting anomalies.
By doing so we are creating a pareto front of implementations, which allows for choice of the best combination of preprocessing, detection and hardware.

\section{Preliminary Data and Conclusion}

In the previous section, we presented our holistic concept for finding the optimal anomaly detection implementation.
The first step to conducting the evaluation is the implementation of a modular feature extraction unit, which can be adjusted to the requirements of any given anomaly detection method.
An overview of our implementation of \emph{Feature Extraction} is given in Figure~\ref{system_overview}.
The design consists of a hash function followed by a buffer for the required parts of network package headers.
From each buffer, metrics such as average, minumum and maximum are computed.
Depending on the configurations of buffer and hash function calculation, different levels of information are stored in the feature extraction unit.
Each network sketch is interpreted as an array with $2^\text{Hash Width}$ entries, while hashes are implemented using shifters.
Furthermore, buffers store memory stages, i.e. varying time periods, to represent the system state.

The implementation is done out of context using Vivado 2023.1 using the Zynq UltraScale+ (XCZU3EG-1SBVA484I).
\begin{table}[!t]
  \caption{Hardware Resources of several design}
  \label{t_measuring}
  \centering
  \begin{tabularx}{\columnwidth}{XX|XXXX}
	  \toprule
    Mem Stages & Hash Width & LUTs & FFs & $f$ (MHz) & Power (W) \\
	\midrule
    1 & 4 & 3.34 \% & 3.28 \% & 444 & 0.410 \\
    3 & 4 & 7.34 \% & 9.48 \% & 430 & 0.733 \\
    1 & 5 & 6.67 \% & 6.20 \% & 471 & 0.617 \\
    3 & 5 & 14.35 \% & 18.21 \% & 465 & 1.295 \\
	\bottomrule
  \end{tabularx}
\end{table}
The results of the system are shown in the Table~\ref{t_measuring}.
It can be seen that saving more information of past generation does to an increase in power and resource consumption.
Furthermore, the network achieves $\qty{430}{\MHz}\cdot\qty{50}{\byte}=\qty{21}{\GBs}$ of throughput.
Therefore, it is able to keep up with the 5G datarates of up to \SI{20}{\GBs}.


\vspace{-2mm}
\renewcommand{\baselinestretch}{0.96}\normalsize
\bibliographystyle{IEEEtran}
\bibliography{hipeac}

\end{document}